\documentclass[pra,twocolumn,superscriptaddress,showpacs]{revtex4}

\usepackage{graphicx,color}
\usepackage{dcolumn}
\usepackage{bm}
\usepackage{ulem}

\usepackage{amsmath}
\usepackage{amssymb,amsbsy}

\newcommand{\beq}{\begin{equation}}
\newcommand{\eeq}{\end{equation}}
\newcommand{\bea}{\begin{eqnarray}}
\newcommand{\eea}{\end{eqnarray}}
\newcommand{\bec}{\begin{center}}
\newcommand{\enc}{\end{center}}
\newcommand{\bfr}{\begin{flushright}}
\newcommand{\efr}{\end{flushright}}

\newcommand{\om}{\omega}
\newcommand{\tom}{\widetilde{\omega}}
\newcommand{\tkap}{\widetilde{\kappa}}

\newcommand{\tone}{\widetilde{1}}
\newcommand{\ttwo}{\widetilde{2}}
\newcommand{\tthree}{\widetilde{3}}
\newcommand{\tfour}{\widetilde{4}}
\newcommand{\tfive}{\widetilde{5}}
\newcommand{\tsix}{\widetilde{6}}
\newcommand{\Om}{\Omega}

\newcommand{\kap}{\kappa}
\newcommand{\gam}{\gamma}

\newcommand{\s}{\sigma}

\newcommand{\la}{\langle}
\newcommand{\ra}{\rangle}
\newcommand{\cH}{{\cal H}}
\newcommand{\cF}{{\cal F}}

\begin{document}
\title{
Dressed-state engineering for continuous detection of itinerant microwave photons 
}
\author{Kazuki Koshino}
\affiliation{College of Liberal Arts and Sciences, Tokyo Medical and Dental
University, Ichikawa, Chiba 272-0827, Japan}
\author{Zhirong Lin}
\affiliation{RIKEN Center for Emergent Matter Science (CEMS), 2-1 Hirosawa, Wako, 
Saitama 351-0198, Japan}
\author{Kunihiro Inomata}
\affiliation{RIKEN Center for Emergent Matter Science (CEMS), 2-1 Hirosawa, Wako, 
Saitama 351-0198, Japan}
\author{Tsuyoshi Yamamoto}
\affiliation{Smart Energy Research Laboratories, NEC Corporation,  
Tsukuba, Ibaraki 305-8501, Japan}
\author{Yasunobu Nakamura}
\affiliation{RIKEN Center for Emergent Matter Science (CEMS), 2-1 Hirosawa, Wako, 
Saitama 351-0198, Japan}
\affiliation{Research Center for Advanced Science and Technology (RCAST), 
The University of Tokyo, Meguro-ku, Tokyo 153-8904, Japan}
\date{\today}
\begin{abstract}
We propose a scheme for continuous detection of itinerant microwave photons
in circuit quantum electrodynamics. 
In the proposed device, a superconducting qubit is coupled dispersively to two resonators: 
one is used to form an impedance-matched $\Lambda$ system 
that deterministically captures incoming photons, 
and the other is used for continuous monitoring of the event. 
The present scheme enables efficient photon detection: 
for realistic system parameters, the detection efficiency reaches 0.9
with a bandwidth of about ten megahertz. 
\end{abstract}
\maketitle

Microwave quantum optics using superconducting qubits and transmission lines,
which is realized in circuit-quantum-electrodynamics setups,  
is one of the hottest research area in modern quantum physics~\cite{cqed2}. 
Exploiting the large dipole moment of superconducting qubits, 
circuit QED enables various quantum-optical phenomena 
that have not been reached by quantum optics in the visible domain. 
In particular, we can readily construct optical setups 
with excellent one-dimensionality~\cite{oned1,oned2,oned3}, 
which are suitable to construct a scalable quantum circuit. 
However, the lack of an efficient detector for itinerant microwave photons 
has been a long-standing problem, 
and several approaches have been proposed to date.  
One approach is to capture a propagating photon deterministically 
into a resonator mode and detect it afterward. 
In recent experiments, the possibility of such capturing 
has been demonstrated with an excellent fidelity~\cite{UCSB1,UCSB2}. 
However, this approach requires precise temporal control of the system parameters
that depends on the exact pulse shapes of the signal photons.
Another approach is to use the Kerr effect  
mediated by superconducting qubits~\cite{kerr1,kerr2,kerr3}, 
which may enable non-destructive photon detection. 
However, it has been revealed that a high distinguishability 
of the signal photon number can be achieved only by 
cascading several identical qubits
with negligible photon loss in between,
which is a challenging technical task presently~\cite{kerr2,kerr3}.

An alternative approach is to use the deterministic switching 
of a $\Lambda$ system induced by individual photons~\cite{KK2009,KK2010,KK2013},
which has been experimentally realized in one-dimensional systems~\cite{Inomata,Dayan}. 
Note that this occurs as a result of single-photon dynamics: 
the destructive interference between the input and the elastically 
scattered photons enables the deterministic operation.
Recently, detection of propagating microwave photons has been demonstrated 
using a $\Lambda$ system realized in a tilted washboard potential 
of a current-biased Josephson junction~\cite{det1,det2,det3}. 
A problem with this scheme could be the substantial dissipation upon detection
and the resultant long dead time before resetting. 
More recently, we realized a $\Lambda$ system formed
by the dressed states of a qubit-resonator system
and discussed its performance as a photon detector~\cite{KK2013,Inomata,KK2015}. 
This detector attains a high detection efficiency within the detection bandwidth, 
regardless of the signal pulse profile and with negligible dark counts. 
However, this detector should be operated in the time-gated mode, 
since the drive field to generate the $\Lambda$-type transition 
must be turned off during the qubit readout to obtain a high fidelity.

\begin{figure}[t]
\includegraphics[width=80mm]{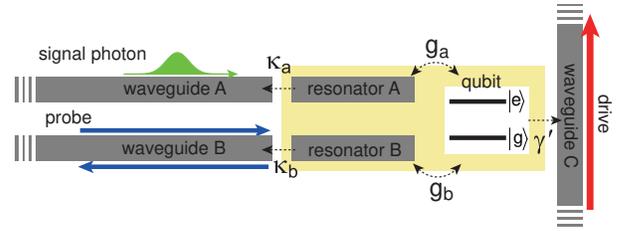}
\caption{\label{fig:sch}
Schematic of the single-photon detector. 
A qubit is coupled dispersively to two resonators.
Resonators~A, B, and the qubit are respectively 
coupled to waveguides~A, B, and C.
We input a signal photon through waveguide~A,
a probe field through waveguide~B,
and a drive field through waveguide~C.
}\end{figure}

In this study, we present a practical scheme 
for continuous detection of itinerant microwave photons. 
We couple two resonators to a qubit: 
one is used for forming a $\Lambda$ system~\cite{KK2013,Inomata} 
and the other is used for continuous qubit monitoring~\cite{read2,read4,read5,read8}. 
The proposed device enables continuous operation of the photon detector, 
preserving the advantages of our previous scheme~\cite{KK2015},
such as a high detection efficiency,  
insensitivity to the signal pulse shape, 
and short dead times after detection. 
Moreover, the efficient detection 
is possible without cascading qubits~\cite{kerr2,kerr3}.

We consider a device in which a superconducting qubit 
is coupled to two resonators~A and B (Fig.~\ref{fig:sch}). 
Setting $\hbar=1$, this system is described by 
$\cH_{sys} = \bar\om_a a^{\dag}a + \bar\om_b b^{\dag}b + \bar\om_q \s^{\dag}\s
+ g_a(a^{\dag}\s+\s^{\dag}a) + g_b(b^{\dag}\s+\s^{\dag}b)$,
where $a$, $b$, and $\s$ respectively denote 
the annihilation operators for resonators~A, B, and the qubit.
$\bar\om_a$, $\bar\om_b$, and $\bar\om_q$ are their bare frequencies,
and $g_a$ and $g_b$ are the qubit-resonator couplings. 
We set $(\bar\om_a, \bar\om_b, \bar\om_q, g_a, g_b)
/2\pi=(10, 12, 5, 0.5, 0.4)$~GHz for concreteness. 
Since this system is in the dispersive regime,
$\cH_{sys}$ is rewritten as 
\bea
\cH_{sys} = (\om_a a^{\dag}a + \om_b b^{\dag}b)\s\s^{\dag}
+ [\om_q +(\om_a-2\chi_a)a^{\dag}a
\nonumber
\\
+(\om_b-2\chi_b)b^{\dag}b ]\s^{\dag}\s, 
\hspace{35mm}
\label{eq:Hsys}
\eea
where $\chi_r=g_r^2/(\bar\om_r-\bar\om_q)$ 
is called the dispersive shift $(r=a,b)$, 
and the renormalized frequencies of the resonators and the qubit
are $\om_a=\bar{\om}_a+\chi_a$, $\om_b=\bar{\om}_b+\chi_b$, 
and $\om_q=\bar{\om}_q-\chi_a-\chi_b$. 
Their values are $(\chi_a, \chi_b)/2\pi=(50, 23)$~MHz and 
$(\om_a, \om_b, \om_q)/2\pi=(10.050, 12.023, 4.927)$~GHz.

The qubit and the resonators are respectively coupled to waveguides,
through which we apply three kinds of microwaves (Fig.~\ref{fig:sch}).
Through waveguide~C, we apply a continuous drive field to the qubit 
to generate an ^^ ^^ impedance-matched'' $\Lambda$ system 
by the dressed states of the qubit and resonator~A. 
Through waveguide~A, we input a signal photon to be detected, which
deterministically induces a Raman transition and excites the qubit. 
Through waveguide~B, we apply a continuous probe field 
for the dispersive readout of the qubit state.

\begin{figure}[t]
\includegraphics[width=85mm]{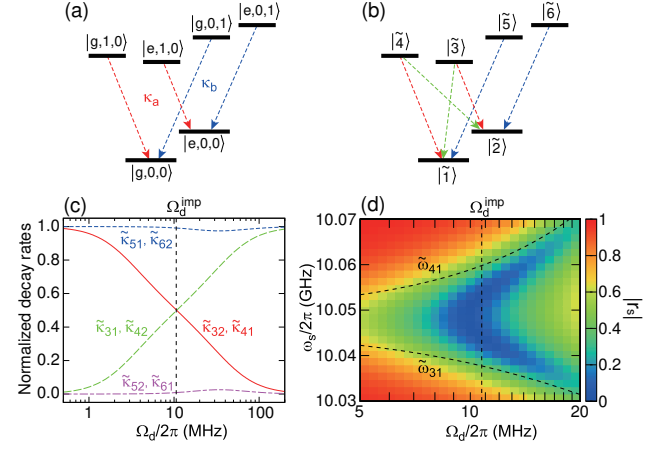}
\caption{\label{fig:imL}
Dressed-state engineering of the qubit-resonators system.
(a)~Level structure of the bare states ($\Om_d=0$) in the rotating frame.
(b)~Level structure of the dressed states
at the operation point ($\Om_d=\Om_d^{imp}$). 
(c)~Dependences of the decay rates on the drive power.
The drive frequency is set at $\om_d/2\pi =4.832$~GHz.
An impedance-matched $\Lambda$ system is formed at $\Om_d^{imp}/2\pi=10.75$~MHz. 
$\tkap^a_{ji}$ ($\tkap^b_{ji}$) is normalized by $\kap_a$ ($\kap_b$).
(d)~Amplitude of the reflection coefficient $|r_s|$ 
of a continuous signal field applied through waveguide~A,
as a function of the signal frequency $\om_s$ and the drive power $\Om_d$.
The upper (lower) curve represents $\tom_{41}$ ($\tom_{31}$).}
\end{figure}

We denote the radiative decay rates of resonators~A, B, and the qubit
by $\kap_a$, $\kap_b$, and $\gam'$, respectively.
$\kap_a$ determines the bandwidth of the photon detector, 
which should be smaller than or comparable to 
the level separation of the dressed states
[$|\tthree\ra$ and $|\tfour\ra$ of Fig.~\ref{fig:imL}(b)]. 
$\kap_b$ determines the phase shift of the probe field upon reflection
and $\kap_b \simeq 2\chi_b$ is favorable for qubit readout~\cite{gam}. 
We set $(\kap_a, \kap_b)/2\pi=(20, 46)$~MHz. 
Additionally, the qubit undergoes non-radiative decay 
and its overall decay rate $\gam$ 
often dominates $\gam'$. 
The photon detection efficiency is sensitive to $\gamma$. 
We assume a reasonably long-lived qubit of $\gam/2\pi=0.01$~MHz 
($T_1 \approx 16~\mu{\rm s}$)~\cite{long1,long2}. 

We engineer the dressed states of qubit-resonators system through the qubit drive. 
Theoretically, the qubit drive is described by 
$\cH_{dr}(t)=\sqrt{\gam'}[E_d(t)\s^{\dag}+E_d^*(t)\s]$,
where $E_d(t)=E_d e^{-i\om_d t}$ is a monochromatic drive field.
In the frame rotating at $\om_d$, we obtain a static Hamiltonian,
\bea
\cH_{sys+dr} = (\om_a a^{\dag}a + \om_b b^{\dag}b)\s\s^{\dag}
+ [(\om_q-\om_d) 
\hspace{15mm}
\nonumber
\\
+(\om_a-2\chi_a)a^{\dag}a
+(\om_b-2\chi_b)b^{\dag}b ]\s^{\dag}\s
+\Om_d(\s^{\dag}+\s),
\label{eq:Hsd}
\eea
where $\Om_d=\sqrt{\gam'}E_d$ is the Rabi frequency of the qubit drive.
Hereafter, $\Om_d$ represents the drive power.

First, we consider the case with $\Om_d=0$.
The eigenstates of $\cH_{sys+dr}$ are the Fock states $|q,n_a,n_b\ra$,
where $q(=g,e)$ denotes the qubit state and 
$n_a$ and $n_b(=0,1,\cdots)$ denote the resonator photon numbers.
To find the optimal drive conditions, we restrict ourselves to
the zero- and one-photon states [Fig.~\ref{fig:imL}(a)]. 
In this study, we use resonator~A to form a $\Lambda$ system
and resonator~B as a readout resonator
that preserves the qubit state upon transitions. 
For this purpose, we should realize the level structure of Fig.~\ref{fig:imL}(a),
where $\om_{|g,0,0\ra}<\om_{|e,0,0\ra}<\om_{|e,1,0\ra}<\om_{|g,1,0\ra}$
(nesting regime for resonator~A) and 
$\om_{|g,0,0\ra}<\om_{|e,0,0\ra}<\om_{|g,0,1\ra}<\om_{|e,0,1\ra}$
(un-nesting regime for resonator~B)~\cite{KK2013}.
This is done by setting 
the drive frequency within the range of $\om_q-2\chi_a < \om_d < \om_q-2\chi_b$.

Next, we consider the case with $\Om_d>0$.
The drive field mixes the bare states to form the dressed states. 
We denote the dressed states by $|\tone\ra, |\ttwo\ra, \cdots$
and label them from the lowest in energy [Fig.~\ref{fig:imL}(b)]. 
The states $|\tone\ra$ and $|\ttwo\ra$ are made of the zero-photon states.
From Eq.~(\ref{eq:Hsd}), 
they are given by
\bea
|\tone\ra &=& \cos\theta_{12}|g,0,0\ra - \sin\theta_{12}|e,0,0\ra,\label{eq:tone}
\\
|\ttwo\ra &=& \sin\theta_{12}|g,0,0\ra + \cos\theta_{12}|e,0,0\ra,
\\
\theta_{12} &=& \arctan[2\Om_d/(\om_{|e,0,0\ra}-\om_{|g,0,0\ra})]/2.\label{eq:theta12}
\eea
$|\tthree\ra$ and $|\tfour\ra$ ($|\tfive\ra$ and $|\tsix\ra$) 
are made of the one-photon states of resonator~A (B), which are 
obtained by replacing $|g,0,0\ra$ and $|e,0,0\ra$
in Eqs.~(\ref{eq:tone})--(\ref{eq:theta12})
with $|e,1,0\ra$ and $|g,1,0\ra$ ($|g,0,1\ra$ and $|e,0,1\ra$).
The radiative decay rate from $|\widetilde{j}\ra$ ($j=3,4$)
to $|\widetilde{i}\ra$ ($i=1,2$) is given by
$\tkap^a_{ji}=\kap_a|\la\widetilde{j}|a^{\dag}|\widetilde{i}\ra|^2$.
We confirm that 
$\tkap^a_{31}=\tkap^a_{42}$, $\tkap^a_{32}=\tkap^a_{41}$, 
and $\tkap^a_{31}+\tkap^a_{32}=\tkap^a_{41}+\tkap^a_{42}=\kap_a$.
Namely, $|\tthree\ra$ and $|\tfour\ra$ decay in two directions, 
satisfying the sum rule of decay rates. 
Similarly, 
$\tkap^b_{51}=\tkap^b_{62}$, $\tkap^b_{52}=\tkap^b_{61}$, 
and $\tkap^b_{51}+\tkap^b_{52}=\tkap^b_{61}+\tkap^b_{62}=\kap_b$. 
Figure~\ref{fig:imL}(c) plots $\tkap^a_{ji}$ and $\tkap^b_{ji}$
as functions of the drive power. 
In the drive-off limit ($\Om_d \to 0$),
$\tkap^a_{32}=\tkap^a_{41} \to \kap_a$, 
$\tkap^b_{51}=\tkap^b_{62} \to \kap_b$, and others vanish.
This represents the simple decay of the resonator modes
preserving the qubit state [Fig.~\ref{fig:imL}(a)].  
As we increase the drive power, 
the decay rates for resonator~A are inverted, 
whereas those for resonator~B remain almost unchanged.
This is because of our choice of the drive frequency $\om_d$. 
At $\Om_d^{imp}$ in Fig.~\ref{fig:imL}(c), 
the four decay rates concerning resonator~A become identical.
Then, 
a resonant signal photon deterministically induces a Raman transition of 
$|\tone\ra \to |\widetilde{j}\ra \to |\ttwo\ra$ ($j=3,4$). 
Regarding resonator~B, 
we should make $\tkap^b_{52}(=\tkap^b_{61})$ as small as possible
to suppress 
the $|\tone\ra \to |\widetilde{k}\ra \to |\ttwo\ra$ and 
$|\ttwo\ra \to |\widetilde{k}\ra \to |\tone\ra$ transitions ($k=5,6$).
For this purpose, 
$\om_d$ close to $\om_q-2\chi_a$ is advantageous.
We set $\om_d/2\pi=4.832$~GHz 
[5~MHz above $(\om_q-2\chi_a)/2\pi$] hereafter,
which results in $\Om_d^{imp}/2\pi=10.75$~MHz. 
Then, $\cos^2\theta_{12}=0.99$, $\cos^2\theta_{34}=0.61$, 
and $\cos^2\theta_{56}=0.96$. 
Namely, the dressed states 
$|\tone\ra$, $|\ttwo\ra$, $|\tfive\ra$, and $|\tsix\ra$
are almost identical to the bare states
$|g,0,0\ra$, $|e,0,0\ra$, $|g,0,1\ra$, and $|e,0,1\ra$, respectively.
$\tkap^b_{52}(=\tkap^b_{61})$ is about 0.9\% of $\kap_b$. 

\begin{figure}[t]
\includegraphics[width=85mm]{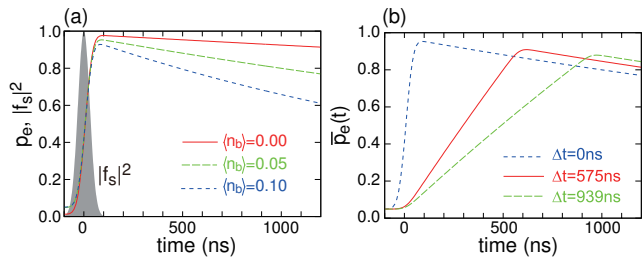}
\caption{\label{fig:tev}
Microwave response to a signal photon.
The signal photon has a carrier frequency of $\om_s/2\pi=10.05$~GHz 
and a Gaussian pulse shape with the length of $l=100$~ns.
(a)~Time evolution of the qubit excitation probability $p_e(t)$.
The probe power $\la n_b \ra$ is indicated. 
The signal photon profile $|f_s(t)|^2$ is also shown 
in units of $(8\ln 2/\pi l^2)^{1/2}$. 
(b)~Time-averaged qubit excitation probability $\overline{p}_e(t)$ for $\la n_b \ra=0.05$. 
The integration time $\Delta t$ is indicated. 
}\end{figure}

In Fig.~\ref{fig:imL}(d), we plot the reflection coefficient $|r_s|$ of 
a weak continuous field of frequency $\om_s$ through waveguide~A. 
We observe that impedance matching ($|r_s|\simeq 0$) takes place
at $\Om_d \simeq \Om_d^{imp}$ and $\tom_{31} \lesssim \om_s \lesssim \tom_{41}$, 
where $\tom_{ji}$ denotes the transition frequency 
between $|\widetilde{j}\ra$ and $|\widetilde{i}\ra$. 
This indicates that each signal photon induces the Raman transition 
in the $\Lambda$ system and is absorbed deterministically.
We choose these drive power and signal frequency 
as the operating points of the photon detector.

Next, we study the response of the detector to a single-photon signal. 
The signal photon is assumed to be a Gaussian pulse  
with length $l$ and frequency $\om_s$, namely,
$f_s(t)=(8\ln 2/\pi l^2)^{1/4} \times 2^{-t^2/(l/2)^2}e^{-i\om_s t}$,
which is normalized as $\int dt |f_s(t)|^2=1$. 
Setting $\om_s/2\pi=10.05$~GHz and $l=100$~ns, 
we plot the time evolution of the qubit excitation probability $p_e(t)$,
which represents the population of $|\ttwo\ra$, 
by the red solid line in Fig.~\ref{fig:tev}(a).
$p_e(t)$ increases within the pulse duration and approaches to unity, 
which agrees well with $\int_{-\infty}^t dt'|f_s(t')|^2$. 
A high efficiency is attained regardless of the pulse shape
as long as the linewidth of the photon 
is narrower than that of the $\Lambda$ system. 
After the pulse duration, 
$p_e(t)$ decreases gradually by natural decay of the qubit with rate $\gamma$. 

\begin{figure*}[t]
\includegraphics[width=140mm]{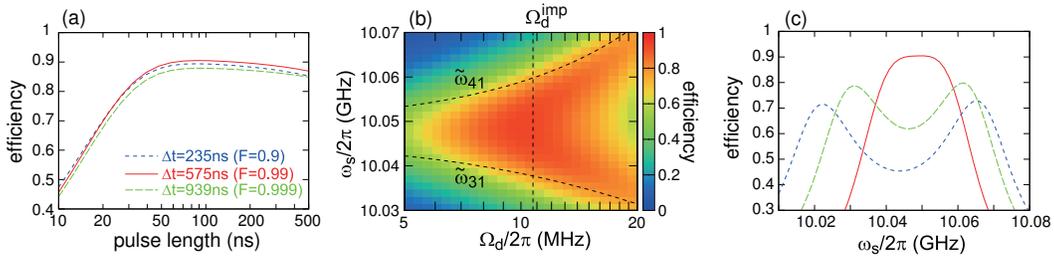}
\caption{\label{fig:eff}
Single-photon detection efficiency. 
The probe power is fixed at $\la n_b \ra=0.05$. 
(a)~Dependence of the detection efficiency on the pulse length $l$
for various integration times $\Delta t$. 
Values of the corresponding readout fidelity $\cF$ are also indicated.
The input photon is tuned at $\om_s/2\pi=10.05$~GHz.
(b)~Detection efficiency as a function of $\Om_d$ and $\om_s$
for $l=100$~ns and $\Delta t=575$~ns.
The dashed lines indicate $\tom_{41}$, $\tom_{31}$, and $\Om_d^{imp}$.
(c)~Cross section of (b) at $\Om_d^{imp}/2\pi=10.75$~MHz (red solid).
The results for different drive conditions are also shown: 
$\om_d/2\pi=4.841$~GHz and $\Om_d^{imp}/2\pi=17.27$~MHz (green dashed) and 
$\om_d/2\pi=4.850$~GHz and $\Om_d^{imp}/2\pi=21.00$~MHz (blue dotted). 
}\end{figure*}
Now we consider the effect of the probe field, $E_p(t)=E_p e^{-i\om_p t}$,
by which the qubit is continuously monitored~\cite{read2,read4,read5,read8}. 
From Eq.~(\ref{eq:Hsys}), the resonant frequency of resonator~B depends on the qubit state. 
This is reflected in the phase shift of the probe field upon reflection. 
The phase shift $\theta_g$ for the qubit ground state is given by
$\theta_g = 2 \arctan[\kap_b/2(\om_b-\om_p)]$, 
and $\theta_e$ for the qubit excited state is
obtained by replacing $\om_b$ with $\om_b-2\chi_b$. 
Hereafter, in order to suppress the $|\tone\ra \to |\tfive\ra$ transition,
we set the probe frequency at $\om_p=\om_b-2\chi_b$
rather than the usually chosen condition $\om_p=\om_b-\chi_b$.  
This results in $e^{i\theta_g}=(3+4i)/5$ and $e^{i\theta_e}=-1$. 
We measure one quadrature of the reflected field 
for discriminating the qubit state. 
We infer the qubit state through the time-averaged probe field 
with an integration time $\Delta t$.
The signal-to-noise ratio (SNR) is given, 
assuming that the noise is purely of quantum origin, by
\beq
\mathrm{SNR}=\sqrt{\frac{\kap_b \la n_b \ra \Delta t}{4}} \left|e^{i\theta_g}-e^{i\theta_e}\right|,
\label{eq:SNR}
\eeq
where 
$\la n_b \ra=4|E_p|^2/\kap_b$ represents the probe power 
in terms of the mean photon number in resonator~B~\cite{read2}. 
The readout fidelity ${\cal F}$ is given by
\beq
{\cal F} = \mathrm{erf}(\mathrm{SNR}/\sqrt{2}),
\label{eq:fid}
\eeq
where erf denotes the error function~\cite{fid}. 
In practice, the noise could be enhanced 
by technical reasons such as the noise from the amplifiers.
Here, we assume the noiseless phase-sensitive amplification
preserving the SNR~\cite{caves}.

In Fig.~\ref{fig:tev}(a), we plot the time evolution of $p_e(t)$
in the presence of the probe field with various power $\la n_b \ra$. 
We observe that the near-deterministic qubit excitation 
[red solid line in Fig.~\ref{fig:tev}(a)]
is degraded by increasing the probe power. 
This is attributed mainly to the 
enhanced qubit decay through
the $|\ttwo\ra \to |\widetilde{k}\ra \to |\tone\ra$ transition ($k=5,6$). 
However, for the probe power considered here, 
the backaction of the qubit readout is not severe 
and the qubit excitation is maintained for several microseconds. 
Hereafter, we fix the probe power at $\la n_b \ra=0.05$.
Then, $\cF=0.99$ (0.999) [SNR=2.58 (3.29)]
is attained by taking $\Delta t=$575 (939)~ns.
The long qubit lifetime enables us to take such long integration times.

The single-photon detection efficiency $\eta$,
which is the probability to find the qubit excitation,
is evaluated as follows. 
Since we infer the qubit state through the time-averaged amplitude,
we introduce the time-averaged excitation probability, 
\beq
\overline{p}_e(t)=\frac{1}{\Delta t}\int_{t-\Delta t}^{t}dt' p_e(t'),
\eeq
and find the maximum probability $\overline{p}_e(t_m)$ in time. 
Considering that the probability to correctly infer the qubit state is $(1+\cF)/2$, 
$\eta$ is given by 
\beq
\eta=\overline{p}_e(t_m)\frac{1+\cF}{2}
+[1-\overline{p}_e(t_m)]\frac{1-\cF}{2}.
\eeq
$\eta \simeq 1/2$ for a low SNR,
implying that the qubit state is completely indistinguishable. 
In contrast, $\eta \simeq \overline{p}_e(t_m)$ for a high SNR.

In Fig.~\ref{fig:tev}(b), we plot $\overline{p}_e(t)$ for various $\Delta t$. 
Expectedly, $\overline{p}_e(t)$ becomes flatter as we increase $\Delta t$,
which implies the loss of detection signal. 
However, owing to the long qubit lifetime, 
the decrease of $\overline{p}_e(t)$ due to the time-averaging 
is at most several percent.
In Fig.~\ref{fig:eff}(a), we plot the efficiency 
as a function of the pulse length $l$ of the signal photon. 
If the qubit lifetime is infinite,  
the efficiency increases monotonically with $l$. 
In practice, the efficiency is maximized at a finite pulse length
due to the qubit decay during the pulse duration.  
We observe that a high efficiency is maintained for a wide range of $l$, 
which is an advantage of the $\Lambda$-based scheme.
The loss of efficiency is due to the infidelity of the qubit measurement for short $\Delta t$,
whereas it is due to the time-averaging for long $\Delta t$.
For $\Delta t=575$~ns ($\cF=0.99$),
the maximum efficiency of 0.91 is obtained at $l\simeq 90$~ns.
In Fig.~\ref{fig:eff}(b), we plot the efficiency as a function of $\Om_d$ and $\om_s$. 
Comparing this with Fig.~\ref{fig:imL}(d), 
we confirm that the impedance-matching leads to a high detection efficiency. 
The cross section of Fig.~\ref{fig:eff}(b) at $\Om_d=\Om_d^{imp}$ 
is shown in Fig.~\ref{fig:eff}(c) by the red solid line,
which shows the detection band of this detector.
The detection efficiency exceeds 0.9 (0.8) for a bandwidth of 9~MHz (20~MHz).

Four final comments are in order.
(i)~The detector is insensitive to signal photons 
when the qubit is excited.
This causes a dead time of the detector,
which amounts to several microseconds at 
$\la n_b\ra=0.05$ [Fig.~\ref{fig:tev}(a)].
However, by applying a reset pulse upon detection of the qubit excitation~\cite{KK2015},
we may shorten the dead time to several hundreds of nanoseconds.
(ii)~The detection band center is tunable by changing the drive condition. 
We show the detection band for different drive conditions in Fig.~\ref{fig:eff}(c). 
The detection band has two peaks located at $\tom_{31}$ and $\tom_{41}$ in general. 
However, as we increase $\om_d$ and accordingly $\Om_d^{imp}$, 
$\tkap_{52}$ and $\tkap_{61}$ are increased.
This enhances the probe backaction and degrades the detection efficiency.
(iii)~In the continuous measurement, one may worry that 
the quantum Zeno effect prohibits efficient photon detection,
since the apparent measurement time interval seems infinitely small. 
However, even in the continuous measurement,
the effective measurement time interval $\Delta t_m$ remains finite,
which is determined by the dephasing rate induced by the measuring apparatus~\cite{zeno1,zeno2}.
Here, the probe field functions as the apparatus and 
$\Delta t_m$ is determined by SNR$\sim$1, 
namely, $\Delta t_m \sim 1/\kap_b \la n_b \ra = 175$~ns.
This is obviously long enough to avoid the Zeno effect 
[see Fig.~\ref{fig:tev}(a)].
(iv)~The probe photons may cause the dark counts by inducing 
the $|\tone\ra \to |\widetilde{k}\ra \to |\ttwo\ra$ transition ($k=5,6$).
We can numerically check that this probability is about 0.2\% per one probe photon. 
Therefore, the dark count rate is estimated to be $(142~\mu{\rm s})^{-1}$ for $\la n_b \ra=0.05$.
A lower dark count rate is accomplished by reducing the probe power.

In summary, we proposed a practical scheme 
for continuous detection of itinerant microwave photons. 
The detector consists of a qubit and two resonators in the dispersive regime. 
We apply a drive field to the qubit to form an impedance-matched $\Lambda$ system, 
a signal photon to one of the resonators to excite the qubit, 
and a probe field to the other to continuously monitor the qubit. 
For realistic system parameters, 
the detector has a maximum detection efficiency exceeding 0.9
and a bandwidth of about ten megahertz. 
One can improve the performance of the detector further
by increasing the qubit lifetime and/or the dispersive shifts. 

This work was partly supported by 
MEXT KAKENHI (Grant Nos. 25400417 and 26220601),
Project for Developing Innovation Systems of MEXT, 
National Institute of Information and Communications Technology (NICT),
and ImPACT Program of Council for Science, Technology and Innovation.

\appendix
\section{Detection efficiency}
In the main part of this study, we defined the detection efficiency intuitively 
through the time-averaged qubit excitation probability. 
Here, we investigate the detection efficiency more rigorously 
on the basis of the quantum jumps of the qubit.
We observe that the deviation between these two definitions 
is negligible for the parameter range discussed in this study. 

\subsection{Time-independent case}
In the dispersive readout of the qubit state, 
we infer the qubit state through the time-averaged probe field. 
First, we preliminarily observe a case in which 
the qubit keeps staying in its ground/excited state. 
We denote the field operator for the probe port by $c(t)$, 
which is normalized as $[c(t), c^{\dagger}(t')]=\delta(t-t')$. 
We apply a classical field (coherent state) as the probe of the qubit state. 
The phase of the probe field is sensitive to the qubit state as
\beq
\la c(t) \ra =  
\begin{cases}
E_p e^{i\theta_g} & \mathrm{for}\ |g\ra
\\
E_p e^{i\theta_e} & \mathrm{for}\ |e\ra
\end{cases}, 
\eeq
where the natural phase factor $e^{-i\om_p t}$ is neglected. 
We introduce the time-averaged field operator by
\beq
\bar{c}(t) = \frac{1}{\sqrt{\Delta t}}\int_{t-\Delta t}^t dt' c(t'), 
\eeq
which is normalized as $[\bar{c}(t), \bar{c}^{\dagger}(t)]=1$. 
We infer the qubit state through one of its quadratures, 
$\bar{x}(t)=\mathrm{Im}[e^{-i(\theta_g+\theta_e)/2}\bar{c}(t)]$,
which maximizes the signal-to-noise ratio (SNR). 
The expectation value of this operator is 
\beq
\la\bar{x}(t)\ra = 
\begin{cases}
-E_p\sqrt{\Delta t}\sin\left(\frac{\theta_e-\theta_g}{2}\right)  & \mathrm{for}\ |g\ra
\\
E_p\sqrt{\Delta t}\sin\left(\frac{\theta_e-\theta_g}{2}\right)  & \mathrm{for}\ |e\ra
\end{cases}. 
\eeq
We set the threshold at $\la\bar{x}\ra=0$ and
judge the qubit state through the sign of $\la\bar{x}\ra$.
Since $\bar{c}(t)$ is normalized as $[\bar{c}(t), \bar{c}^{\dag}(t)]=1$, 
the quantum noise in each quadrature is 1/2 for a coherent state. 
The SNR and the readout fidelity are then given by
\bea
\mathrm{SNR} &=& 2E_p\sqrt{\Delta t}\sin\left(\frac{\theta_e-\theta_g}{2}\right),
\\
\cF &=& \mathrm{erf}(\mathrm{SNR}/\sqrt{2}),
\eea
which are Eqs.~(\ref{eq:SNR}) and (\ref{eq:fid}) of the main part. 
The probability to correctly infer the qubit state is $(1+\cF)/2$. 

\subsection{Time-dependent case}
Next we investigate a more realistic situation in which 
the qubit is excited at $t \sim 0$ and decays gradually afterward [Fig.~3(a)]. 
The detection efficiency $\eta$ is defined 
as the probability to detect the qubit excitation. 
We compare two methods for evaluating this probability:
In method~1, which we adopted in the main part of this study, 
we intuitively evaluate the detection efficiency $\eta_1$ 
through the time-averaged qubit excitation probability.
In method~2, we evaluate the detection efficiency $\eta_2$ more rigorously 
based on the quantum jumps of the qubit observed in actual measurements. 

\subsubsection{Method~1}
In the main part of this study, 
we evaluate the detection efficiency as follows. 
From the qubit excitation probability $p_e(t)$, 
we define the time-averaged probability $\bar{p}_e(t)$ by
\beq
\bar{p}_e(t) = \frac{1}{\Delta t}\int_{t-\Delta t}^t dt' p_e(t'), 
\eeq
and find the moment $t_m$ that maximizes $\bar{p}_e(t)$. 
We define the detection efficiency $\eta_1$ 
as the probability to detect the qubit excitation at this moment. 
Considering that the probability to correctly infer the qubit state is $(1+\cF)/2$, 
$\eta_1$ is given by
\beq
\eta_1 = \bar{p}_e(t_m)\frac{1+\cF}{2}+[1-\bar{p}_e(t_m)]\frac{1-\cF}{2}.
\label{eq:eta1}
\eeq

\subsubsection{Method~2}
The qubit excitation/de-excitation is observed as the quantum jumps in actual measurements. 
We consider a single event where the qubit is excited at $t=0$ and is de-excited at $t=\tau$. 
Considering the rapid response time of the resonator 
($1/\kappa_b \simeq 3.5$~ns), we may regard that 
the probe field responds immediately to the quantum jumps of the qubit as
\beq
\la c(t) \ra = 
\begin{cases}
E_p e^{i\theta_g} & (t<0, \tau<t)
\\
E_p e^{i\theta_e} & (0 \leq t \leq \tau)
\end{cases}. 
\eeq
$\la\bar{x}\ra$ is maximized at $t=(\tau+\Delta t)/2$. 
The maximum value depends 
on the duration $\tau$ of the qubit excitation as
\beq
\la\bar{x}\ra = 
\begin{cases}
-E_p\sqrt{\Delta t}\sin\left(\frac{\theta_e-\theta_g}{2}\right)
\left(1-2\frac{\tau}{\Delta t}\right)  & (0<\tau<\Delta t)
\\
E_p\sqrt{\Delta t}\sin\left(\frac{\theta_e-\theta_g}{2}\right)  & (\Delta t<\tau)
\end{cases}. 
\eeq
Accordingly, the probability $q(\tau)$ to detect the qubit excitation is
\beq
q(\tau) = 
\begin{cases}
\frac{1}{2} \left\{1-\mathrm{erf} \left[
\frac{\mathrm{SNR}}{\sqrt{2}} \left(1-2\frac{\tau}{\Delta t}\right)
\right]\right\} & (0 \leq \tau \leq \Delta t)
\\
\frac{1}{2} \left[1+\mathrm{erf} \left(\frac{\mathrm{SNR}}{\sqrt{2}} 
\right)\right]  & (\Delta t<\tau)
\end{cases}. 
\eeq
The shape of $q(\tau)$ is shown in Fig.~\ref{fig:S}(a). 
It is a monotonically increasing function of $\tau$ 
and becomes constant for $\tau>\Delta t$.

We denote the probability distribution function of the duration $\tau$ 
of the qubit excitation by $Q(\tau)$, 
which is normalized as $\int_0^{\infty}d\tau Q(\tau)=1$. 
Then, the overall probability $\eta_2$ to detect the qubit excitation is
\beq
\eta_2 = \int_0^{\infty}d\tau Q(\tau)q(\tau). 
\eeq

\subsubsection{Comparison of $\eta_1$ and $\eta_2$}
\begin{figure}[t]
\includegraphics[width=85mm]{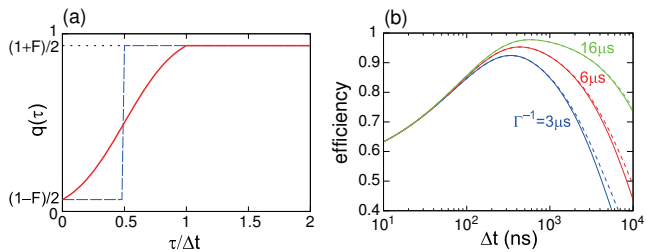}
\caption{\label{fig:S}
(a)~The probability $q(\tau)$ to detect the qubit excitation
as a function of the excitation duration time $\tau$ (red solid),
and its step-function approximation (blue dashed).
(b)~Comparison of $\eta_1$ (dotted) and $\eta_2$ (solid). 
The qubit lifetime $\Gamma^{-1}$ is assumed to be
3~$\mu$s (blue), 6~$\mu$s (red), and 16~$\mu$s (green).  
}\end{figure}
Here we compare $\eta_1$ and $\eta_2$ assuming a simple case 
in which the qubit is excited at $t=0$ and decays exponentially with a rate of $\Gamma$. 
The excitation probability $p_e(t)$ is given by
\beq
p_e(t) = \theta(t) e^{-\Gamma t},
\eeq
where $\theta(t)$ is the step function. 
The probability distribution $Q(\tau)$ is connected to $p_e(t)$ by
$\int_t^{\infty}d\tau Q(\tau)=p_e(t)$. $Q(\tau)$ is therefore given by
\beq
Q(\tau) = \Gamma e^{-\Gamma\tau}. 
\eeq

We can show that $\eta_1$ and $\eta_2$ are almost identical
if the qubit decay within the integration time $\Delta t$ is small,
namely, $\Gamma \Delta t \lesssim 1$. 
Regarding $\eta_1$, $\bar{p}_e(t)$ is maximized at $t_m=\Delta t$, 
and the maximum value $\bar{p}_e(t_m)$ is approximated well by $p_e(\Delta t/2)$.
Equation~(\ref{eq:eta1}) is then rewritten as
\beq
\eta_1 = p_e(\Delta t/2) \frac{1+\cF}{2}+[1-p_e(\Delta t/2)]\frac{1-\cF}{2}.
\label{eq:eta1b}
\eeq
On the other hand, regarding $\eta_2$, 
we may replace $q(\tau)$ by a step function [dashed line in Fig.~\ref{fig:S}(a)]
as long as $Q(\tau)$ is almost constant for $0 \leq \tau \leq \Delta t$. 
Then, using 
$\int_{\Delta t/2}^{\infty} d\tau Q(\tau)=p_e(\Delta t/2)$, 
$\eta_2$ is rewritten as
\beq
\eta_2=p_e(\Delta t/2)\frac{1+\cF}{2}+[1-p_e(\Delta t/2)]\frac{1-\cF}{2},
\eeq
which coincides with $\eta_1$ of Eq.~(\ref{eq:eta1b}).

In Fig.~\ref{fig:S}(b), $\eta_1$ and $\eta_2$ are plotted 
as functions of the integration time $\Delta t$, 
setting the probe power at $\la n_b \ra=0.05$. 
We confirm that the deviation between $\eta_1$ and $\eta_2$ is small
for $\Delta t \lesssim \Gamma^{-1}$. 
From Fig.~\ref{fig:tev}(a), we estimate that 
the qubit lifetime is about 6~$\mu$s for $\la n_b \ra=0.05$. 
Then, $|\eta_1-\eta_2|\lesssim 0.01\%$ for $\Delta t \lesssim 1$~$\mu$s. 
Thus, we can safely regard $\eta_1$ as the detection efficiency. 



\begin{thebibliography}{99}
\bibitem{cqed2}
A. Wallraff, D. I. Schuster, A. Blais, L. Frunzio, R. S. Huang, J. Majer, S. Kumar, S. M. Girvin, 
and R. J. Schoelkopf,
Nature {\bf 431}, 162 (2004).

\bibitem{oned1}
O. Astafiev, A. M. Zagoskin, A. A. Abdumalikov Jr., Yu. A. Pashkin,
T. Yamamoto, K. Inomata, Y. Nakamura, and J. S. Tsai, 
Science {\bf 327}, 840 (2010).
\bibitem{oned2}
I.-C. Hoi, C. M. Wilson, G. Johansson, T. Palomaki, 
B. Peropadre, and P. Delsing, 
Phys. Rev. Lett. {\bf 107}, 073601 (2011).
\bibitem{oned3}
A. F. van Loo, A. Fedorov, K. Lalumiere, B. C. Sanders,
A. Blais, and A. Wallraff, Science {\bf 342}, 1494 (2013).


\bibitem{UCSB1}
Y. Yin, Y. Chen, D. Sank, P. J. J. O'Malley, 
T. C. White, R. Barends, J. Kelly, E. Lucero, 
M. Mariantoni, A. Megrant, C. Neill, A. Vainsencher, 
J. Wenner, A. N. Korotkov, A. N. Cleland, and J. M. Martinis,
Phys. Rev. Lett. {\bf 110}, 107001 (2013).
\bibitem{UCSB2}
J. Wenner, Y. Yin, Y. Chen, R. Barends, B. Chiaro, 
E. Jeffrey, J. Kelly, A. Megrant, J. Y. Mutus, 
C. Neill, P. J. J. O'Malley, P. Roushan, D. Sank, 
A. Vainsencher, T. C. White, A. N. Korotkov, A. N. Cleland, 
and J. M. Martinis,
Phys. Rev. Lett. {\bf 112}, 210501 (2014).

\bibitem{kerr1}
F. Helmer, M. Mariantoni, E. Solano, and F. Marquardt, 
Phys. Rev. A {\bf 79}, 052115 (2009).
\bibitem{kerr2}
S. R. Sathyamoorthy, L. Tornberg, A. F. Kockum, B. Q. Baragiola, J. Combes, 
C. M. Wilson, T. M. Stace, and G. Johansson,
Phys. Rev. Lett. {\bf 112}, 093601 (2014). 
\bibitem{kerr3}
B. Fan, G. Johansson, J. Combes, G. J. Milburn, and T. M. Stace, 
Phys. Rev. B {\bf 90}, 035132 (2014).
\bibitem{KK2009}
K. Koshino,
Phys. Rev. A {\bf 79}, 013804 (2009). 
\bibitem{KK2010}
K. Koshino, S. Ishizaka, and Y. Nakamura,
Phys. Rev. A {\bf 82}, 010301(R) (2010). 
\bibitem{KK2013}
K. Koshino, K. Inomata, T. Yamamoto, and Y. Nakamura,
Phys. Rev. Lett. {\bf 111}, 153601 (2013).

\bibitem{Inomata}
K. Inomata, K. Koshino, Z. R. Lin, W. D. Oliver, J. S. Tsai, Y. Nakamura, and T. Yamamoto,
Phys. Rev. Lett. {\bf 113}, 063064 (2014).
\bibitem{Dayan}
I. Shomroni, S. Rosenblum, Y. Lovsky, O. Bechler, G. Guendelman, and B. Dayan, 
Science {\bf 345}, 903 (2014).

\bibitem{det1}
Y.-F. Chen, D. Hover, S. Sendelbach, L. Maurer, S. T. Merkel, 
E. J. Pritchett, F. K. Wilhelm, and R. McDermott,
Phys. Rev. Lett. {\bf 107}, 217401 (2011).
\bibitem{det2}
B. Peropadre, G. Romero, G. Johansson, C. M. Wilson, E. Solano, and J. J. Garc\'{i}a-Ripoll,
Phys. Rev. A {\bf 84}, 063834 (2011).
\bibitem{det3}
A. Poudel, R. McDermott, and M. G. Vavilov,
Phys. Rev. B {\bf 86}, 174506 (2012).

\bibitem{KK2015}
K. Koshino, K. Inomata, Z. Lin, Y. Nakamura, and T. Yamamoto,
Phys. Rev. A {\bf 91}, 043805 (2015).

\bibitem{read2}
R. Vijay, D. H. Slichter, and I. Siddiqi, 
Phys. Rev. Lett. {\bf 106}, 110502 (2011).

\bibitem{read4}
M. Hatridge, S. Shankar, M. Mirrahimi, F. Schackert, K. Geerlings,
T. Brecht, K. M. Sliwa, B. Abdo, L. Frunzio, S. M. Girvin,
R. J. Schoelkopf, and M. H. Devoret, 
Science {\bf 339}, 178 (2013). 
%
\bibitem{read5}
Z. R. Lin, K. Inomata, W. D. Oliver, K. Koshino, Y. Nakamura, J. S. Tsai, and T. Yamamoto,
Appl. Phys. Lett. {\bf 103} 132602 (2013). 
%
\bibitem{read8}
B. Abdo, K. Sliwa, S. Shankar, M. Hatridge, L. Frunzio, R. Schoelkopf, and M. Devoret, 
Phys. Rev. Lett. {\bf 112}, 167701 (2014).

\bibitem{gam}
J. Gambetta, A. Blais, M. Boissonneault, A. A. Houck, D. I. Schuster, and S. M. Girvin, 
Phys. Rev. A {\bf 77}, 012112 (2008).

\bibitem{long1}
J. Bylander, S. Gustavsson, F. Yan,	F. Yoshihara, K. Harrabi, G. Fitch, D. G. Cory,
Y. Nakamura, J.-S. Tsai, and W. D. Oliver, 
Nature Phys. {\bf 7}, 565 (2011). 

\bibitem{long2}
C. Rigetti, J. M. Gambetta, S. Poletto, B. L. T. Plourde, J. M. Chow, A. D. Corcoles, J. A. Smolin, 
S. T. Merkel, J. R. Rozen, G. A. Keefe, M. B. Rothwell, M. B. Ketchen, and M. Steffen, 
Phys. Rev. B {\bf 86}, 100506(R) (2012).

\bibitem{fid}
J. Gambetta, W. A. Braff, A. Wallraff, S. M. Girvin, and R. J. Schoelkopf,
Phys. Rev. A {\bf 76}, 012325 (2007). 
\bibitem{caves}
C. M. Caves, Phys. Rev. D {\bf 26}, 1817 (1982).
\bibitem{zeno1}
L. S. Schulman, Phys. Rev. A {\bf 57}, 1509 (1998).
\bibitem{zeno2}
K. Koshino and A. Shimizu, 
Phys. Rep. {\bf 412} 191 (2005).

\end{thebibliography}
\end{document}